\RequirePackage{fix-cm}
\documentclass[smallextended]{svjour3}       
\smartqed  
\usepackage{graphicx}
\usepackage{float}

\newcommand{\cG}{{\cal G}}
\newcommand{\cF}{{\cal F}}

\newcommand{\beq}{\begin{equation}}
\newcommand{\eeq}{\end{equation}}

\newcommand{\ber}{\begin{eqnarray}} 
\newcommand{\eer}{\end{eqnarray}}
\def\bf{\textbf}

\begin{document}

\title{Two-photon exchange correction in elastic lepton-proton scattering
}

\author{Oleksandr Tomalak}


\institute{O. Tomalak \at
              Johann-Joachim-Becher-Weg 45\\
              D55128 Mainz, Germany \\
              Tel.: +4917697695451\\
              \email{tomalak@uni-mainz.de} 
}

\date{Received: date / Accepted: date}

\maketitle

\begin{abstract}
We present the dispersion relation approach based on unitarity and analyticity to evaluate the two-photon exchange contribution to elastic electron-proton scattering. The leading elastic and first inelastic $\pi N$ intermediate state contributions are accounted for in the region of small momentum transfer $Q^2 < 1~\mathrm{GeV}^2$ based on the available data input. The novel methods of analytical continuation allow us to exploit the MAMI form factor data and the MAID parameterization for the pion electroproduction amplitudes as input in the calculation. The results are compared to the recent CLAS, VEPP-3 and OLYMPUS data as well as to the full two-photon exchange correction in the near-forward approximation, which is based on the Christy and Bosted unpolarized structure functions fit. Additionally, predictions are given for a forthcoming muon-proton scattering experiment.

\end{abstract}

\vspace{-0.3cm}
\section{Introduction}
\label{intro}

Two-photon exchange (TPE) corrections to elastic electron-proton scattering are the leading unknown contributions in the analysis of the experimental elastic lepton-proton data. According to the studies of the A1 Collaboration \cite{Bernauer:2010wm,Bernauer:2013tpr}, these corrections could be useful for the extraction of the proton magnetic radius and elastic form factors from electron-proton scattering data. Model-independent determination of radii and form factors are required as input to the evaluation of the TPE correction to hyperfine splitting \cite{Carlson:2008ke,Carlson:2011af,Tomalak:2017npu,Tomalak:2017lxo}. It is also of growing importance for the realization and analysis of the forthcoming measurements of the ground state hyperfine splitting in muonic hydrogen with 1 $\mathrm{ppm}$ accuracy level by CREMA~\cite{Pohl:2016tqq}, FAMU~\cite{Adamczak:2016pdb} collaborations and at J-PARC~\cite{Ma:2016etb}.

We evaluate the TPE contribution to the unpolarized scattering cross section at small scattering angles approximating the hadronic part of the TPE graph as an unpolarized forward Compton scattering process. We present the data-driven dispersion relation approach and evaluate the elastic and pion-nucleon TPE contributions within this framework. Additionally, we provide the first estimates of the TPE correction at the kinematics of the forthcoming muon-proton scattering experiment accounting for all mass terms.

\section{Elastic lepton-proton scattering and two-photon exchange}
\label{sec"0}

\begin{figure}[h]
\begin{center}
\includegraphics[width=.35\textwidth]{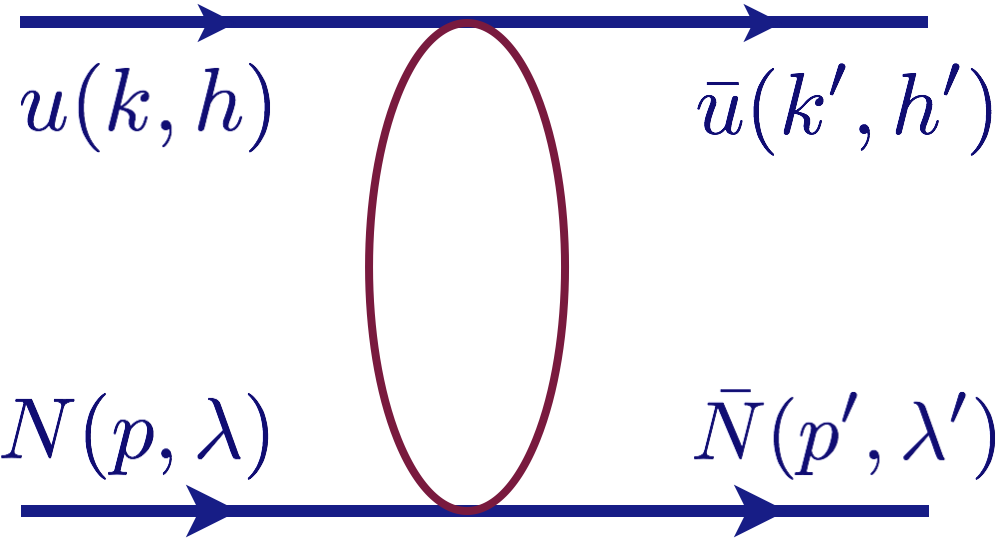}
\end{center}
\caption{Elastic lepton-proton scattering.}
\label{elastic_scattering_general}
\end{figure}

Elastic lepton-proton scattering $ l( k , h ) + p( p, \lambda ) \to l( k^\prime, h^\prime) + p(p^\prime, \lambda^\prime) $, as in Fig. \ref{elastic_scattering_general}, where we indicate the momenta $k,p$ ($k^\prime,p^\prime$) and helicities $h,\lambda$ ($h^\prime,\lambda^\prime$) of incoming (outgoing) particles, is completely described by 2 Mandelstam variables, e.g., $ Q^2 = - (k-k^\prime)^2 $ - the squared momentum transfer, and $ s = ( p + k )^2 $ - the squared energy in the lepton-proton center-of-mass reference frame.
In a dispersion relation analysis, it is convenient to introduce the crossing symmetric variable $\nu$: $ \nu = (s-u)/4 $ which changes sign with $ s\leftrightarrow u $ channel crossing; $u$ denotes the $u$-channel squared energy: $ u = ( k - p^\prime )^2 $. In elastic electron-proton scattering experiments, a convenient variable is  the virtual photon polarization parameter $ \varepsilon $:
\ber \label{epsilon_def}
\varepsilon = \frac{16 \nu^2 - Q^2 ( Q^2 + 4M^2 )}{16 \nu^2 - Q^2 ( Q^2 + 4M^2 ) + 2 ( Q^2 + 4M^2 )( Q^2 - 2 m^2)},
\eer
where $M$ and $m$ are the masses of proton and lepton respectively.

The helicity amplitude $ T_{h^\prime \lambda^\prime, h \lambda} $ for $ l^{-} p $ elastic scattering can be divided into a part without the flip of lepton helicity, and a part with lepton helicity flip $ T^{\mathrm{flip}} $, which is proportional to the mass of the lepton \cite{Goldberger:1957ac,Gorchtein:2004ac} (the $T$ matrix is defined as $ S = 1 + i ~T $):
\ber \label{str_ampl} 
T_{h^\prime \lambda^\prime, h \lambda}^{\mathrm{non-flip}} & = & \frac{e^2}{Q^2} \bar{u}(k^\prime,h^\prime) \gamma_\mu u(k,h) \nonumber \\ 
&& \bar{N}(p^\prime,\lambda^\prime) \left( \gamma^\mu \cG_M (\nu, Q^2) -   \frac{P^{\mu}}{M} \cF_2 (\nu, Q^2) \right) N(p,\lambda) \nonumber \\
& + & \frac{e^2}{Q^2} \cF_3 (\nu, Q^2) \bar{u}(k^\prime,h^\prime) \gamma_\mu u(k,h) \cdot \bar{N}(p^\prime,\lambda^\prime)  \frac{\gamma . K P^{\mu}}{M^2}  N(p,\lambda) , \label{str_ampl1} \\
 T_{h^\prime \lambda^\prime, h \lambda}^{\mathrm{flip}} & = &\frac{e^2}{Q^2} \frac{m}{M} \bar{u}(k^\prime,h^\prime) u(k,h)  \nonumber \\ && \bar{N}(p^\prime,\lambda^\prime)\left( \cF_4 (\nu, Q^2)  +  \frac{\gamma . K}{M} \cF_5 (\nu, Q^2) \right) N(p,\lambda)  \nonumber \\
& + & \frac{e^2}{Q^2} \frac{m}{M} \cF_6 (\nu, Q^2) \bar{u}(k^\prime,h^\prime) \gamma_5 u(k,h) \cdot \bar{N}(p^\prime,\lambda^\prime) \gamma_5 N(p,\lambda), \label{str_ampl2}
\eer
with the averaged momentum variables $ P = (p+p^\prime)/2, ~K = (k+k^\prime)/2 $ and the unit of electric charge $e$. In the $1\gamma$-exchange approximation, only the amplitudes $\cG_M$ and $\cF_2$ present. They are expressed in terms of the proton electric $G_E$ and magnetic $G_M$ form factors as
\ber
\cG^{1\gamma}_M &=& G_M, \\
\cF^{1\gamma}_2 &=& \frac{G_M - G_E}{1+\tau_P},
\eer
with  $\tau_P = Q^2/(4 M^2)$ (see Refs. \cite{Gorchtein:2004ac,Tomalak:2014dja} for a full description of terms). 

The TPE correction at the leading $ \alpha $ order, $ \delta_{2 \gamma} $, is defined through the ratio between the cross section with account of the exchange of two photons and the cross section in the $1 \gamma $-exchange approximation $ \sigma_{1 \gamma} $ by 
\ber \label{TPE_definition}
 \sigma = \sigma_{1 \gamma} \left( 1 + \delta_{2 \gamma} \right).
\eer
The leading TPE correction to unpolarized elastic $ l^{-} p $ scattering can be expressed in terms of the TPE invariant amplitudes as
\ber \label{delta_TPE_massive}
\delta_{2\gamma}   =  \frac{2}{ G_M^2 + \frac{\varepsilon}{\tau_P} G_E^2} \left\{ G_M \Re \cG^{2\gamma}_1 + \frac{\varepsilon}{\tau_P} G_E \Re \cG^{2\gamma}_2 \right. \nonumber \\
\left. + \frac{ 1 - \varepsilon }{ 1 - \varepsilon_0 } \left( \frac{\varepsilon_0}{\tau_P}  \frac{\nu}{M^2}  G_E \Re \cG^{2\gamma}_4 - G_M \Re \cG^{2\gamma}_3 \right) \right\}, 
\eer
with $ \varepsilon_0 = 2m^2/Q^2 $ and the following amplitudes:
\ber  \label{amplitudes_G1}
 \cG^{2\gamma}_1 & = & \cG^{2\gamma}_M + \frac{\nu}{M^2} \cF^{2\gamma}_3 + \frac{m^2}{M^2} \cF^{2\gamma}_5,  \\\label{amplitudes_G2}
 \cG^{2\gamma}_2 & = & \cG^{2\gamma}_M - ( 1 + \tau_P ) \cF^{2\gamma}_2 + \frac{\nu}{M^2} \cF^{2\gamma}_3,  \\\label{amplitudes_G3}
 \cG^{2\gamma}_3 & = & \frac{m^2}{M^2} \cF^{2\gamma}_5 + \frac{\nu}{M^2} \cF^{2\gamma}_3,   \\\label{amplitudes_G4}
 \cG^{2\gamma}_4 & = &  \cF^{2\gamma}_4 + \frac{\nu}{M^2 (1+\tau_P)} \cF^{2\gamma}_5 .
\eer

In this work, we exploit the Maximon and Tjon prescription \cite{Maximon:2000hm} for the infrared-divergent part of the TPE contribution, subtracting the following infrared-divergent term $\delta^{\mathrm{IR}}_{2 \gamma}$ \cite{Tomalak:2014dja}:
\ber \label{IR_delta_TPE}
\delta^{\mathrm{IR}}_{2 \gamma} & =&  \frac{2 \alpha}{\pi}  \ln \left(\frac{Q^2}{\mu^2}\right) \left\{  \frac{s - M^2 - m^2}{\sqrt{\Sigma_s}}   \ln \left(\frac{\sqrt{\Sigma_s}-s+(M+m)^2}{\sqrt{\Sigma_s}+s-(M+m)^2}\right) \right. \nonumber \\  
&& \left.  -  \frac{u - M^2 - m^2}{\sqrt{\Sigma_u}}  \ln \left(\frac{\sqrt{\Sigma_u}-u+(M+m)^2}{ - \sqrt{\Sigma_u} -u + (M+m)^2}\right) \right \},
\eer
with $ \Sigma_s = (s-(M +m)^2)(s-(M -m)^2) $, $ \Sigma_u = (u-(M +m)^2)(u-(M -m)^2) $ and a small photon mass $ \mu $, which regulates the infrared divergence.

\section{Near-forward calculation}
\label{sec:1}

At relatively small lepton scattering angles, we account for all inelastic intermediate states by generalizing the calculation in forward kinematics \cite{Tomalak:2015aoa}. The hadronic part of the TPE graph is approximated as near-forward unpolarized doubly-virtual Compton scattering. Contracting it with the lepton line, we reproduce the leading terms in the momentum transfer expansion \cite{Brown:1970te}:
\ber \label{low_Q2}
\delta_{2 \gamma} \approx a~\sqrt{Q^2} + b~Q^2 \ln Q^2 + c~Q^2 \ln^2 Q^2 + O(Q^2)
\eer
The leading $\sqrt{Q^2}$ term comes from the classical Feshbach result \cite{McKinley:1948zz}, which corresponds to the scattering of the relativistic charged particle in the Coulomb field. The proton intermediate state contributes to all terms in Eq. (\ref{low_Q2}), while inelastic states contribute only to $ Q^2 \ln Q^2 $ term \cite{Brown:1970te}.

On top of the proton state contribution (Born TPE) \cite{Blunden:2003sp}, we express the contribution from the unpolarized proton structure functions $F_1$ and $F_2$ as the weighted integral over the invariant mass of the intermediate state $ W^2 $ and the averaged virtuality of two photons $\tilde{Q}^2$:
\ber \label{TPE_expression}
\delta^{\mathrm{F_1,F_2}}_{2 \gamma} (\nu,~Q^2) = \int \mathrm{d} W^2    \mathrm{d} \tilde{Q}^2 && \left \{ w_1\left(W^2,\tilde{Q}^2,\nu,Q^2\right) F_1\left(W^2, \tilde{Q^2} \right) \right. \nonumber \\
&& \left. + ~w_2\left(W^2,\tilde{Q}^2,\nu,Q^2\right) F_2 \left(W^2, \tilde{Q^2}\right) \right \},
\eer
with the weighting functions $w_1$ and $w_2$. We use the empirical fit performed by Christy and Bosted (BC) \cite{Christy:2007ve} for the numerical evaluation.

In the following Fig. \ref{near_forward}, we compare the total TPE as a sum of the Born TPE and inelastic contributions of Eq. (\ref{TPE_expression}) with the Born TPE only (box diagram model in Fig. \ref{near_forward}), the Feshbach result, and with the empirical TPE fit performed by the MAMI/A1 Collaboration \cite{Bernauer:2013tpr}. The Feshbach correction corresponds to the point-like proton. The Born TPE, which accounts for the distribution of the charge and magnetization inside the proton, has larger TPE at low $\varepsilon$ and smaller TPE at large $\varepsilon$. The account of the inelastic excitations returns the correction close to the Feshbach result at large $\varepsilon$. The total TPE is in a reasonable agreement with the empirical fit of the MAMI/A1 Collaboration \cite{Bernauer:2013tpr}.
\begin{figure}[H]
  \includegraphics[width=\textwidth]{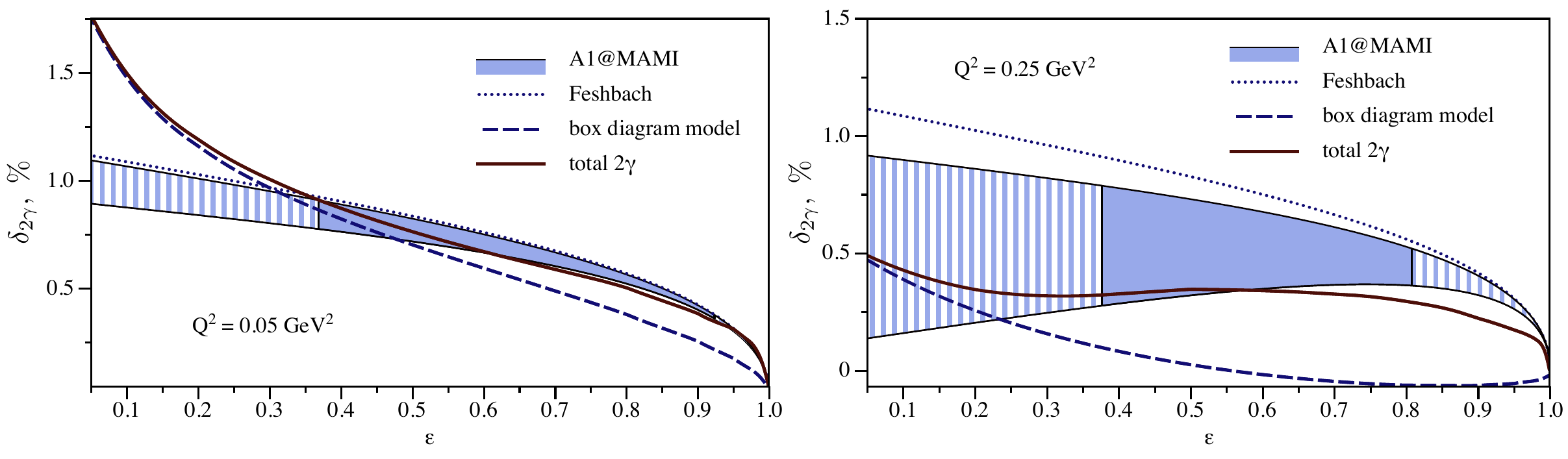}
\caption{$ \varepsilon$ dependence of the TPE correction $ \delta_{2 \gamma} $ to $ e^{-} p \to e^{-} p $ for the fixed momentum transfers $ Q^2 = 0.05 ~\mathrm{GeV}^2 $ (left panel) and $ Q^2 = 0.25 ~\mathrm{GeV}^2 $ (right panel). The Feshbach term for point-like particles, the box graph model evaluation (Born TPE) with dipole form factors, and the total TPE correction as the sum of Born TPE and inelastic TPE are presented. The experimental input for the proton structure functions is taken from the Christy-Bosted fit \cite{Christy:2007ve}. The correction is compared with the empirical TPE fit of Ref. \cite{Bernauer:2013tpr} (A1 Collaboration). The filled region is covered by data.}
\label{near_forward}       
\end{figure}

\section{Fixed-$Q^2$ dispersion relation approach}
\label{sec:2}

At arbitrary scattering angles, the dispersion relation (DR) approach allows us to evaluate the TPE correction as a sum of the contributions from each intermediate state. We realize the DR approach for the fixed value of the momentum transfer and account for the elastic and $\pi N$ intermediate states.

Unitarity relations allow us to relate the imaginary parts of TPE amplitudes at the leading order in $ \alpha $ to the experimental input in a model-independent way. The imaginary part of the TPE helicity amplitude $ \Im T^{2 \gamma}_{h^\prime \lambda^\prime,h \lambda}$ can be evaluated by the phase-space integration of the product of the one-photon exchange amplitudes from initial to intermediate state $ T^{1 \gamma}_{ \mathrm{hel} , h \lambda}  $ and from the intermediate state to final state $ T^{1 \gamma}_{h^\prime \lambda^\prime, \mathrm{hel}} $:
\ber \label{unitarity_relation}
\Im T^{2 \gamma}_{h^\prime \lambda^\prime,h \lambda} & = & \frac{1}{2} \sum \limits_{n,\mathrm{hel} } \prod \limits_{i=1}^n \int \frac{\mathrm{d}^3 \bf{q}_i}{(2 \pi)^3} \frac{1}{2 E_i} ( T^{1 \gamma}_{\mathrm{hel}  , h^\prime \lambda^\prime} )^{*} T^{1 \gamma}_{ \mathrm{hel} , h \lambda} (2 \pi)^4 \delta^4 (k+p-\sum_i q_i), \nonumber \\
\eer
where $ q_i = ( E_i, \bf{q}_i ) $ denotes the momentum of an intermediate particle and the sum goes over all possible number of particles $ n $ and all possible helicity states (denoted as "$ \mathrm{hel} $"). The structure amplitudes are given then by the linear combination of the helicity amplitudes. Each multiparticle intermediate state has a corresponding contribution in Eq. (\ref{unitarity_relation}) and can be treated in the DR approach separately.

The TPE amplitudes $ \cG^{2 \gamma}_M(\nu,Q^2), ~\cF^{2 \gamma}_2(\nu,Q^2), ~\cG^{2 \gamma}_1(\nu,Q^2), ~\cG^{2 \gamma}_2(\nu,Q^2) $ are odd functions $ {\cal{G}}^{\mathrm{\mathrm{odd}}}$  under the crossing $ \nu \to - \nu $, whereas the amplitude $ \cF^{2 \gamma}_3$ is even in $\nu$. In the Regge limit $\nu\rightarrow\infty$ and $Q^{2}/\nu \rightarrow 0$, the functions ${\cal G}_{1,2},~{\cal F}_3$ vanish according to the unitarity constraints \cite{Kivel:2012vs}. This allows us to write the unsubtracted DRs for these amplitudes \cite{Borisyuk:2008es,Tomalak:2014sva,Tomalak:2016vbf}:
\ber
 \label{oddDR}
 \Re {\cal{G}}^{\mathrm{\mathrm{odd}}}(\nu, Q^2) & = & \frac{2 \nu}{\pi} \int \limits^{~ \infty}_{\nu_{\mathrm{thr}}} \frac{\Im {\cal{G}}^{\mathrm{\mathrm{odd}}} (\nu^\prime, Q^2)}{{\nu^\prime}^2-\nu^2}  \mathrm{d} \nu^\prime, \\
 \label{evenDR}
 \Re  {\cal{F}}^{2 \gamma}_3 (\nu, Q^2) & = & \frac{2}{\pi} \int \limits^{~ \infty}_{\nu_{\mathrm{thr}}}  \nu^\prime \frac{\Im  {\cal{F}}^{2 \gamma}_3  (\nu^\prime, Q^2)}{{\nu^\prime}^2-\nu^2}  \mathrm{d} \nu^\prime.
\eer
Eqs. (\ref{oddDR}) and (\ref{evenDR}) are evaluated from the $s$-channel threshold upwards. For nonforward scattering, the elastic threshold is always outside the physical region of lepton-proton scattering and input from the unphysical region is required in Eqs. (\ref{oddDR}) and (\ref{evenDR}). For the analytical continuation, we exploit the contour deformation method \cite{Tomalak:2014sva} and perform the calculation with the proton elastic form factors of Ref. \cite{Bernauer:2013tpr}. We describe the analytical continuation in the case of the $\pi N$ intermediate state in the following subsection.

\subsection{Analytical continuation for $\pi N$ intermediate states}
\label{sec:21}

\begin{figure}[H]
  \includegraphics[width=0.975\textwidth]{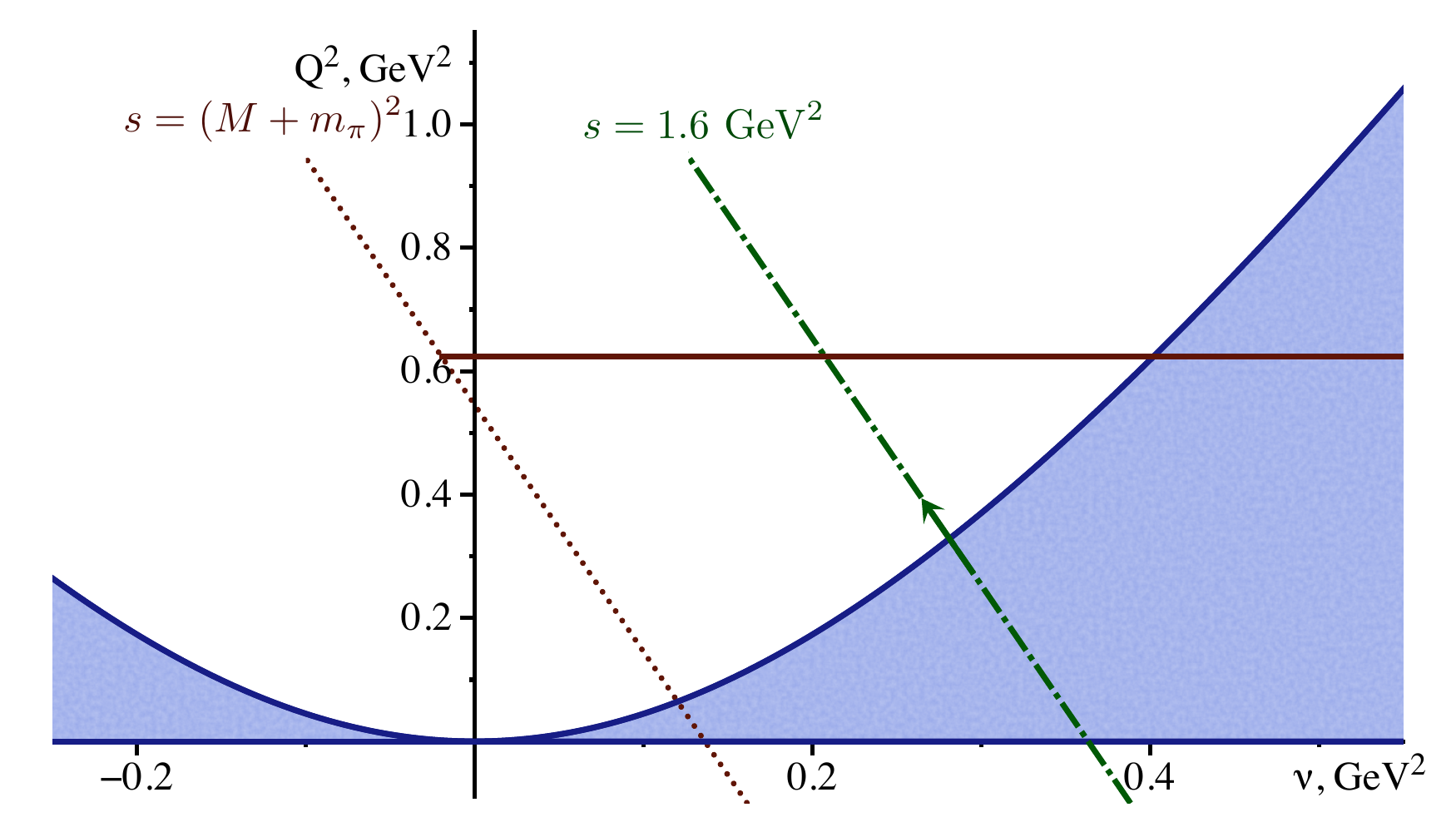}
\caption{Physical and unphysical regions of the kinematical variables $ \nu $ and $ Q^2 $ (Mandelstam plot) for the elastic electron-proton scattering. The hatched blue region corresponds to the physical region, the red-dotted line gives the pion-nucleon ($\pi N$) threshold position in the $s$-channel, the green dashed-dotted line corresponds with the threshold position in the $s$-channel of the state with the invariant mass $ W^2 = 1.6~\mathrm{GeV^2}$ and represents the path of the analytical continuation. The horizontal red curve at fixed $Q^2=0.624~\mathrm{GeV}^2$ illustrates the path of the dispersive integral.}
\label{mandelstam}       
\end{figure}

We illustrate the physical and unphysical regions of the elastic electron-proton scattering and show the pion production threshold in Fig. \ref{mandelstam}. At low momentum transfer $Q^2 \leq 0.624~\mathrm{GeV}^2$, the dispersive integral for $ \pi N$ contribution is evaluated entirely from the physical region. 

\begin{figure}[H]
  \includegraphics[width=\textwidth]{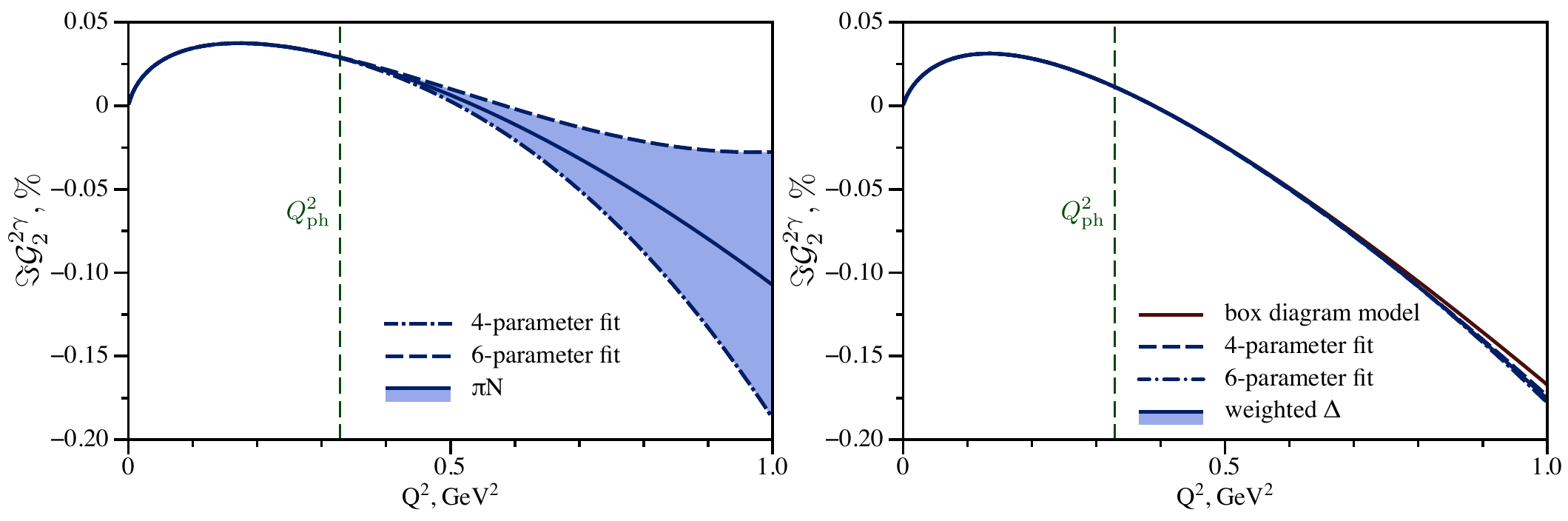}
\caption{The imaginary part of the TPE amplitude $\cG^{2 \gamma}_2$ from the $ \pi N $ (left panel) and weighted-$\Delta$ (right panel) intermediate state contributions as reconstructed from fits of Eq. (\ref{weights}) for the c.m. squared energy $ s = 1.607~\mathrm{GeV}^2$. The analytical continuation of the TPE contribution from the $\Delta$ intermediate state is compared with the exact result in the box diagram model.  The vertical lines correspond with the boundary between the physical ($Q^2 < Q^2_{\mathrm{ph}}$) and unphysical ($Q^2 > Q^2_{\mathrm{ph}}$) regions: $ Q^2_{\mathrm{ph}} \approx 0.329 ~\mathrm{GeV}^2 $.}
\label{imaginary_reconstruction_fits}       
\end{figure}

To calculate the dispersive integral at larger momentum transfer, we perform the analytical continuation for the fixed value of the lepton energy, or $s$, from the physical region at low $Q^2$ to larger $Q^2$. First, we evaluate the imaginary parts in the physical region \cite{Tomalak:2016vbf} exploiting the pion electroproduction amplitudes from the MAID2007 fit~\cite{Drechsel:1998hk,Drechsel:2007if}. Then we fit, for a fixed value of $s$, the $Q^2$ dependence obtained by a sum of the leading terms in the $Q^2$ expansion of the inelastic TPE amplitudes \cite{Brown:1970te,Gorchtein:2014hla,Tomalak:2015aoa,Tomalak_PhD}:
\ber
\Im \cG^{2 \gamma}_1 \left(s,~Q^2 \right) & \sim & Q^2 f \left(s,~Q^2 \right), \label{amplitude_g1} \\
\Im \cG^{2 \gamma}_2 \left(s,~Q^2 \right)  & \sim & Q^2 f \left(s,~Q^2 \right), \\
\Im \cF^{2 \gamma}_3 \left(s,~Q^2 \right)  & \sim & f \left(s,~Q^2 \right),  \label{amplitude_f3}
\eer
with a form for the fitting function:
\ber \label{fitting_function}
f (s,~Q^2) &\equiv & a_1( s ) + a_2( s )  \ln Q^2 + a_3( s )  Q^2 + a_4( s )  Q^2 \ln Q^2 \nonumber \\
 & + & a_5( s )  Q^4 + a_6( s )  Q^4 \ln Q^2.
\eer
We describe the unphysical region by extrapolating the fit of Eqs. (\ref{amplitude_g1})-(\ref{amplitude_f3}). We estimate the theoretical error of the extrapolation procedure as the difference between two fits with four and six parameters, labeled by $f_1$ and $f_2$ respectively. The TPE amplitude $ \Im \cG^{2\gamma}$ is then given by
\ber \label{weights}
\Im \cG^{2\gamma} \left( s,~Q^2 \right)= \frac{ f_1 \left(s,~Q^2 \right) + f_2 \left(s,~Q^2 \right)}{2} \pm \frac{ | f_1 \left(s,~Q^2 \right) - f_2 \left(s,~Q^2 \right)|}{2} ,
\eer
where $f_1$ and $f_2$ have functional forms as in Eq. (\ref{fitting_function}). We illustrate this procedure on the example of the amplitude $ \Im \cG^{2\gamma}_2 $ for the c.m. squared energy $s=1.607~\mathrm{GeV}^2$ in Fig. \ref{imaginary_reconstruction_fits}. For comparison, we also provide the same realization for the test case of the $\Delta$ intermediate state with a finite width, when we know the amplitudes both in physical and unphysical regions. The procedure of the analytical continuation successfully passes the test for the $\Delta$ resonance contribution.

%
%
\subsection{Comparison with data}
\label{sec:22}
\begin{figure}[H]
  \includegraphics[width=\textwidth]{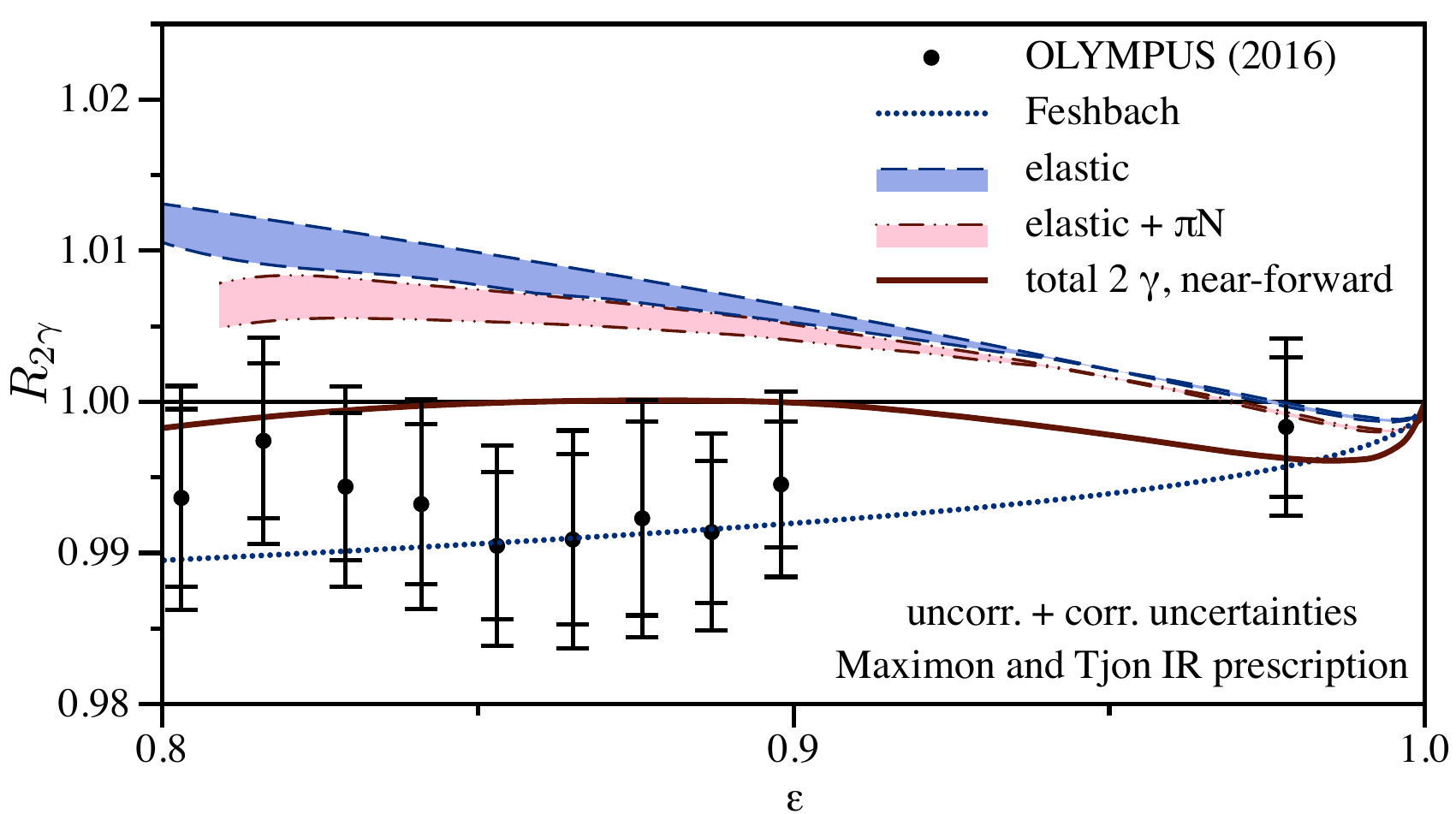}
\caption{The DR result for the elastic TPE and for the sum of elastic and $\pi N$ TPE contributions to the $e^+ p$ over $e^- p$ elastic scattering cross section ratio $ R_{2 \gamma}$ for lepton beam energy $ \omega = 2.01~\mathrm{GeV}$ in comparison with the data from the Olympus Coll. \cite{Henderson:2016dea}. We also show the Feshbach correction \cite{McKinley:1948zz}, as well as the total TPE in the near-forward approximation  of Ref.~\cite{Tomalak:2015aoa}.}
\label{OLYMPUS}       
\end{figure}

\begin{figure}[H]
  \includegraphics[width=\textwidth]{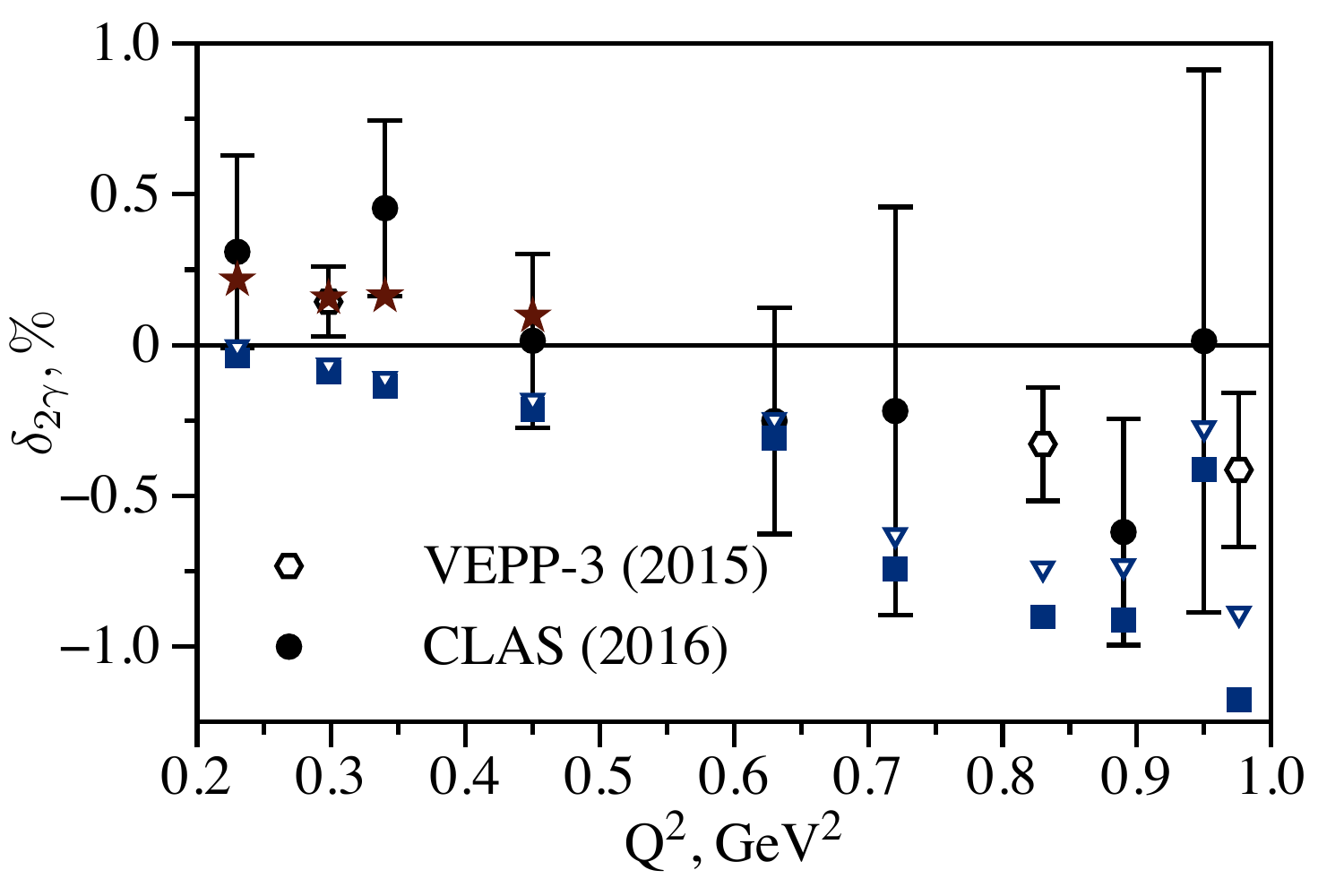}
\caption{TPE correction measurements of Refs.~\cite{Rachek:2014fam,Rimal:2016toz} in comparison with the elastic TPE (shown by squares), and the sum of elastic + $ \pi N$ TPE (shown by hollow triangles). For $Q^2 < 0.5~\mathrm{GeV}^2$, we also compare with the total near-forward TPE of Ref.~\cite{Tomalak:2015aoa} (shown by stars). The VEPP-3 data points were renormalized to the empirical fit of Ref.~\cite{Bernauer:2013tpr} by a procedure which is explained in Ref. \cite{Rachek:2014fam}.}
\label{delta_CLAS_Zhan_088}       
\end{figure}

We provide a comparison of our dispersive and near-forward calculations with the recent data points from the CLAS, VEPP-3, and OLYMPUS experiments \cite{Rachek:2014fam,Henderson:2016dea,Adikaram:2014ykv} in Figs. \ref{OLYMPUS}, \ref{delta_CLAS_Zhan_088} \cite{Tomalak:2017shs}. We present the elastic, the sum of elastic and $\pi N$ TPE contributions as well as the  Feshbach result and the total TPE in the near-forward approximation. OLYMPUS data points at low momentum transfer are accidentally quite well described by the Feshbach correction. The elastic TPE contribution alone is systematically above the data points for $Q^2 > 0.5~\mathrm{GeV}^2$. The data points are described better after an account of the $\pi N$ contribution. However, the $2$-$3\sigma$ difference is still present. The extrapolation of the near-forward calculation allows us to describe the data within $1$-$1.5\sigma$. The near-forward calculation is also in a good agreement with the VEPP-3 and CLAS data points at low $Q^2$. The inclusion of the $\pi N$ TPE contribution improves the description of the CLAS and VEPP-3 data points. However, the CLAS data point at $Q^2=0.35~\mathrm{GeV^2}$ and VEPP-3 data points are more than $1\sigma$ away from the dispersive estimate.
\section{MUSE prediction}
\label{sec:3}

\begin{figure}[H]
  \includegraphics[width=\textwidth]{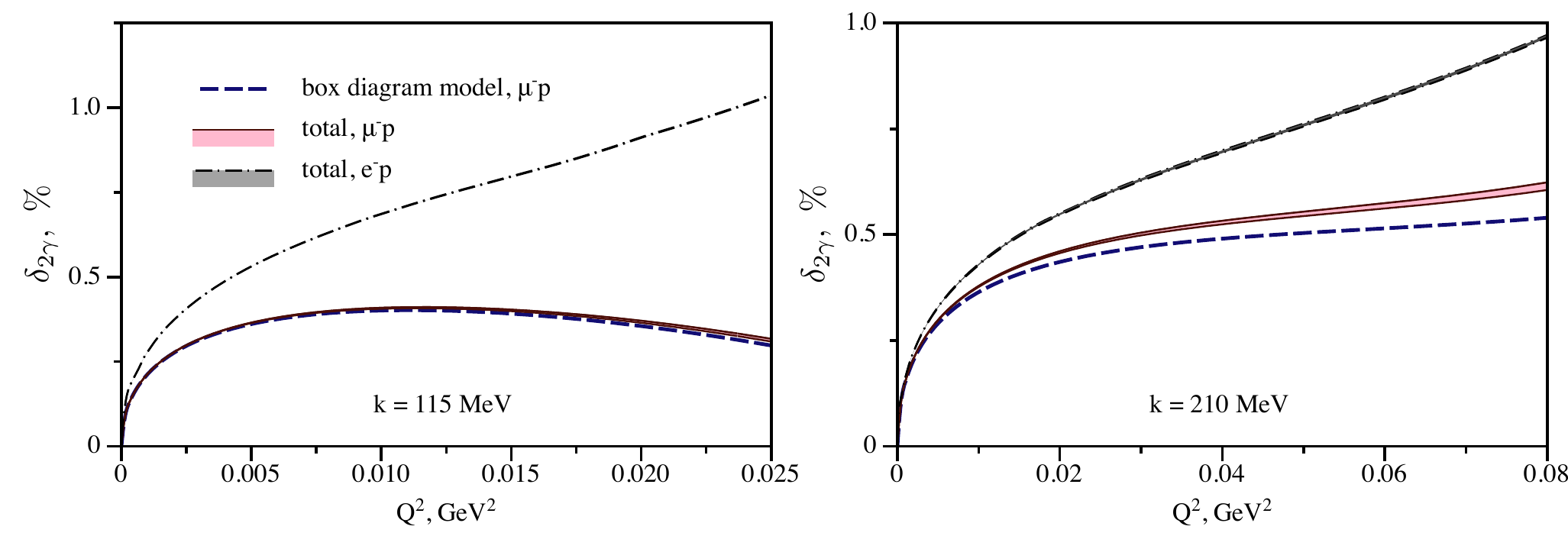}
\caption{The total TPE correction for elastic $\mu^- p$ scattering in kinematics of the MUSE experiment is compared with the Born TPE and the total TPE correction in elastic $ e^- p $ scattering with the same beam momenta.}
\label{MUSE}       
\end{figure}

To shed light on the proton radius puzzle and to extract the charge radius from scattering data with muons, new muon-proton scattering experiment (MUSE) was proposed \cite{Gilman:2013eiv,Gilman:2017hdr}. It aims to simultaneously measure electron, positron, muon and antimuon scattering on a proton target. At low momentum transfer and energies of this experiment, we estimate the TPE correction as a sum of the proton state contribution within the hadronic model and inelastic contributions in the near-forward approximation. We present our estimates in Fig. \ref{MUSE}. The proton state TPE in the hadronic model \cite{Blunden:2003sp} was generalized to the case of massive leptons in Ref. \cite{Tomalak:2014dja}. Due to the cancellation of the helicity-flip and non-flip contributions, the resulting TPE in muon-proton scattering is 2-3 times smaller than the corresponding correction in electron-proton scattering. The inelastic contributions in the near-forward approximation are an order of magnitude below the resulting TPE. The evaluated correction will be useful in the analysis of the forthcoming data from MUSE. The dispersive calculation in the kinematics of this experiment is in progress \cite{Tomalak_PhD}.

\section{Conclusions and outlook}
\label{sec:3}

\begin{figure}[H]
  \includegraphics[width=\textwidth]{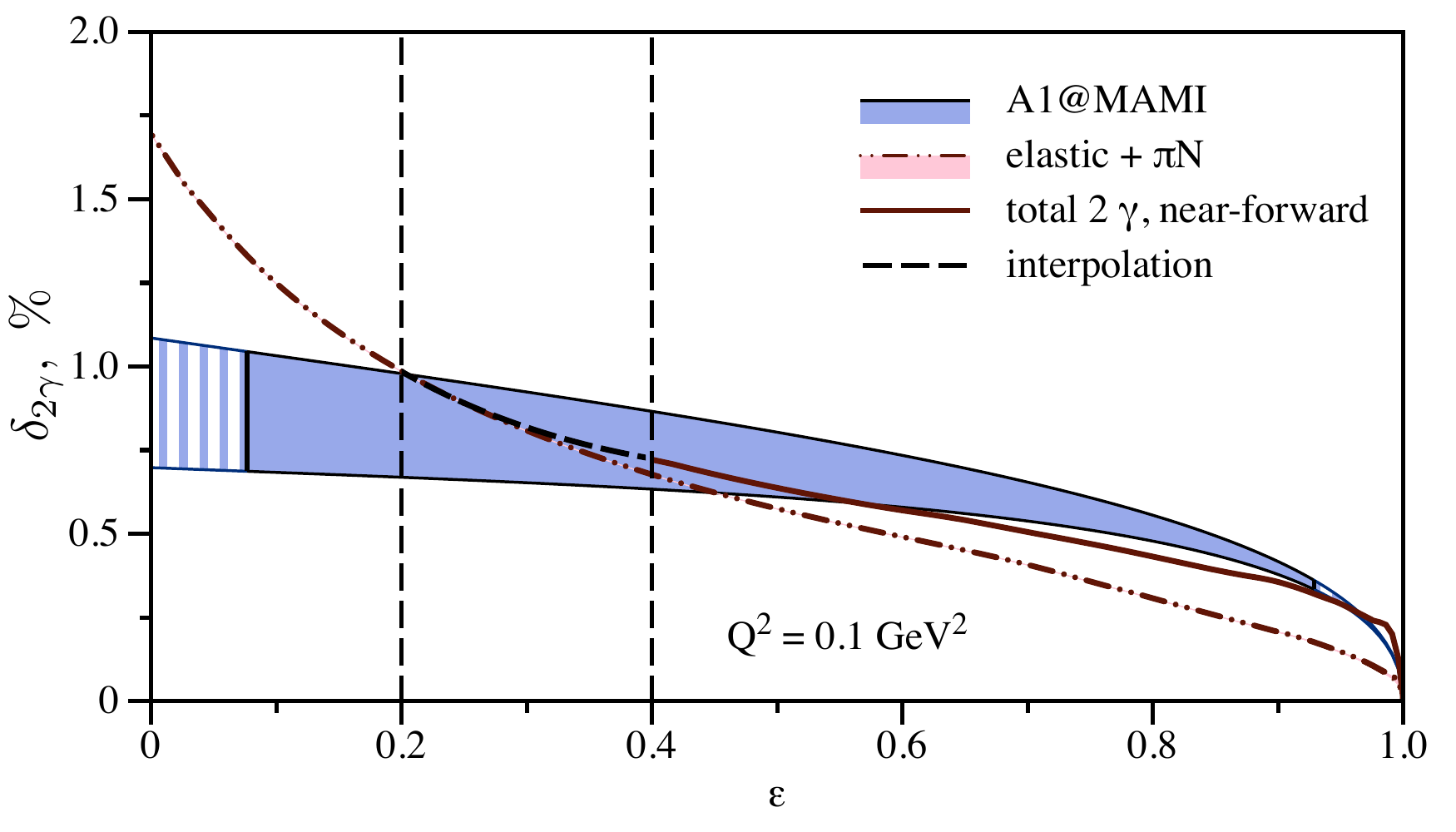}
\caption{TPE correction at low momentum transfer $ Q^2 = 0.1~\mathrm{GeV}^2$. The sum of elastic and $\pi N $ TPE contributions is compared with the total TPE correction in the near-forward approximation and with the empirical TPE fit of Ref. \cite{Bernauer:2013tpr} (A1 Collaboration), where the filled region is covered by data.}
\label{delta_DRQ2_01}       
\end{figure}

The near-forward approximation is in a reasonable agreement with the recent measurements of CLAS, VEPP-3 and OLYMPUS experiments at low momentum transfer. Accounting for the $\pi N$ intermediate state on top of the elastic TPE, the resulting correction comes closer to the experimental data, in comparison to the elastic contribution only, confirming the cancellation between the inelastic TPE and the proton form factor effect, which was previously found in Ref.~\cite{Tomalak:2015aoa}. Our best knowledge of the TPE correction at low momentum transfer is shown in Fig. \ref{delta_DRQ2_01}. 

At small scattering angles (large $\varepsilon$), the near-forward approximation accounts for all inelastic intermediate states. Going to smaller $\varepsilon$, the extrapolation of this calculation is in a good agreement with the empirical extraction of Ref. \cite{Bernauer:2013tpr}. At backward scattering angles, the elastic and pion-nucleon contributions are accounted for within the dispersion relation approach. The intermediate region is described as an interpolation between two calculations. The $\pi N$ TPE correction can be now exploited for precise extraction of the proton magnetic radius and the proton magnetic form factor at low values of $Q^2$.

\begin{acknowledgements}
I thank Marc Vanderhaeghen and Barbara Pasquini for the supervision and support during this work, Carl Carlson for reading this manuscript, Lothar Tiator for useful discussions and providing me with MAID programs, Dalibor Djukanovic for providing me with the access to computer resources. I acknowledge the computing time granted on the supercomputer Mogon at Johannes Gutenberg University Mainz (hpc.uni-mainz.de). This work was supported by the Deutsche Forschungsgemeinschaft DFG in part through the Collaborative Research Center [The Low-Energy Frontier of the Standard Model (SFB 1044)], and in part through the Cluster of Excellence [Precision Physics, Fundamental Interactions and Structure of Matter (PRISMA)].
\end{acknowledgements}



\end{document}